\newcommand{\diracslash}[1]{#1\llap{/\kern2pt}}

\newcommand{\be}{\begin{equation}}
	\newcommand{\ee}{\end{equation}}
\newcommand{\bea}{\begin{eqnarray}}
	\newcommand{\eea}{\end{eqnarray}}
\newcommand{\ba}[1]{\begin{array}{#1}}
	\newcommand{\ea}{\end{array}}

\documentclass[prd,11pt,aps,floats,nofootinbib,tightenlines,showpacs]{revtex4-1}
\usepackage{epsfig,graphicx,pstricks}
\usepackage{psfrag}
\usepackage{color}
\usepackage{amsmath}
\usepackage{mathtools}
\usepackage{amsfonts}
\usepackage{amssymb}
\usepackage{textcomp}
\usepackage{multirow}
\usepackage{subfigure}

\usepackage{stackengine}
\stackMath

\begin{document}
	
	\title {Dynamic density correlations in baryon rich fluid  using Mori-Zwanzig-Nakajima projection operator method}
	\author{Guruprasad Kadam }
	\email{guruprasadkadam18@gmail.com}
	\affiliation{School of Physical Sciences, National Institute of Science Education and Research
		Bhubaneswar, HBNI, Jatni 752050, Odisha, India}

	\date{\today} 
	
	\def\be{\begin{equation}}
		\def\ee{\end{equation}}
	\def\bearr{\begin{eqnarray}}
		\def\eearr{\end{eqnarray}}
	\def\zbf#1{{\bf {#1}}}
	\def\bfm#1{\mbox{\boldmath $#1$}}
	\def\hf{\frac{1}{2}}
	\def\sl{\hspace{-0.15cm}/}
	\def\omit#1{_{\!\rlap{$\scriptscriptstyle \backslash$}
			{\scriptscriptstyle #1}}}
	\def\vec#1{\mathchoice
		{\mbox{\boldmath $#1$}}
		{\mbox{\boldmath $#1$}}
		{\mbox{\boldmath $\scriptstyle #1$}}
		{\mbox{\boldmath $\scriptscriptstyle #1$}}
	}

	\begin{abstract}
In this work, we calculate the dynamic density correlations using Mori-Zwanzig-Nakajima projection operator method. With a judicious choice of slow variables we derive the evolution equations for these slow variables starting from generalised Langevin equation. We get the hydrodynamic form of density correlations function which consist of two acoustic peaks: also called Brillouin peaks, and one thermal peak: also called Rayleigh peak. We then estimate the dynamic density correlations near the critical point using  critical exponents extracted from the statistical bootstrap model of hadronic matter.   We find that the bulk viscosity contributes to the sound attenuation at leading order  $\sim|t|^{-\frac{5}{4}}$ while the thermal conductivity contribute at sub-leading order $\sim|t|^{-\frac{3}{4}}$. On the other hand, only the thermal conductivity contributes at leading order $\sim|t|^{\frac{1}{4}}$ to the thermal diffusivity. We discuss the implications of these results in search for the QCD critical point in the heavy-ion collision experiments.
	\end{abstract}
	
	
	\maketitle

\section{Introduction}

Understanding the quantum chromodynamics (QCD) phase diagram is one of the major challenge in the high energy particle physics and astrophysics today\cite{Fukushima:2010bq,Hands:2001ve,Shuryak:1996pb}. Main impediments in this pursuit are practical and conceptual problems of lattice quantum chromodynamics (LQCD) to understand the phenomenologically relevant part of the phase diagram at finite baryon density. LQCD suffers from so called sign problem at finite chemical potential, however for small chemical potentials there are few reliable results available\cite{Aoki:2006br,Aoki:2009sc,Borsanyi:2010bp,Borsanyi:2020fev}. So one has to resort to various effective models of QCD at relatively high baryon density, $viz.$, Nambu-Jona-Lasinio model (NJL)\cite{Klevansky:1992qe,Hatsuda:1994pi}, Quark-Meson Coupling model(QMC)\cite{Schaefer:2007pw}, Hadron Resonance Gas model (HRG)\cite{Venugopalan:1992hy}, Relativistic Mean-Field model(RMF)\cite{Kadam:2019peo,Pal:2020ucy,Pal:2021qav}, Statistical Bootstrap Model (SBM)\cite{Hagedorn:1965st,Frautschi:1971ij,Satz:1978us} etc.    In the QCD phase diagram a particularly interesting point is the conjectured critical end point (CEP) which is an end point of first order phase transition line.  A lot of theoretical studies has been carried out to find the possible signatures of CEP \cite{Bluhm:2020mpc,Stephanov:2004xs,Dore:2020jye,Mroczek:2020rpm,An:2021wof,Bzdak:2019pkr,Stephanov:2011pb,Grossi:2021gqi,Florio:2021jlx,Datta:2012pj,Gavai:2011zz,Kapusta:2021oco,Kapusta:2012zb} and on the experimental side, the Beam Energy Scan (BES) program has been devoted at the Super Proton Synchrotron (SPS) and at the Relativistic Heavy Ion Collider (RHIC) to search for possible signatures of the CEP \cite{Aggarwal:2010cw, Luo:2017faz}.

  Fluid described by the theory of hydrodynamics belongs to a class of systems in which non-equilibrium dynamics  can be described using only a small number of so-called effective "slow" variables\cite{Landau1987Fluid,Romatschke:2009im,Jeon:2015dfa}. For instance, one can describe the dynamics of a fluid with only few  number of equations of motion: conservation equations for energy density, momentum density, fluid velocity and, possibly, charge density if there are any conserved charges, despite the microscopic degrees of freedom are very large. It turns out that this description can be accommodated into a non-equilibrium transport equation which consist of reversible, irreversible and noise parts. While the former two describes slow dynamics, the noise term contain the information about leftover fast degrees of freedom. Langevin equation is archetypal example of this formalism\cite{zwanzig2001nonequilibrium,balakrishnan2008elements,mazenko2008nonequilibrium}.
 
 A powerful way to study these aspects is Mori-Zwanzig-Nakajima (MZN) projection operator formalism\cite{nakajima1958quantum,mori1965transport,nordholm1975systematic}. In this formalism one can derive the non-equilibrium transport equation by coarse-graining procedure. One first choose a set of slow variables describing long-time dynamics of a system under consideration. Operators corresponding to a set of observables forms a Hilbert space with appropriately defined scalar product. One then  introduces the projection operator which project onto subspace formed by the slow variables.  While the MZN formalism has been traditionally applied to non-relativistic systems\cite{grabert2006projection,te2020projection},  it has recently attracted a lot of attention in high energy physics\cite{Koide:2005qb,Koide:2008nw,Huang:2010sa,Minami:2012hs} and cosmology research\cite{Vrugt:2021sfu}.  
 
 In present work we use the time-independent MZN projection operator method to study the dynamic density correlations in a baryon rich fluid. In the theory of non-equilibrium phenomena correlation functions are very important quantities as they are directly measurable. In the linear response theory the response functions are linearly related to the external perturbations and the time-correlation functions express the propagation of equilibrium fluctuations in the system.  Typically, for a fluid described by hydrodynamics the density correlation function consist of three peaks: two of them corresponds to acoustic waves and one corresponds to thermal fluctuations. It has been noted in previous study\cite{Minami:2009hn} that for a relativistic fluid the width of acoustic peaks, also known as Brillouin peaks, picks up a purely relativistic correction, while the Rayleigh peak due to thermal fluctuations remains unchanged. We shall discuss these aspects in this work.  We shall  further discuss the impact of  critical singularities in the thermodynamic functions and transport coefficients on the Brillouin and Rayleigh peaks. In particular, we shall contrast the critical behaviour of SBM with that of 3D Ising model and compare the results for density correlations as computed in these two models. One of the main advantage of SBM, as we will see, is that one can naturally bring out critical behaviour of hadronic matter near the critical point\cite{Satz:1978us}. Apart from that the exponentially rising density of states has proven absolutely essential to explain the thermodynamics of hadronic matter\cite{Kadam:2018hdo} as well as the small viscosity near $T_c$\cite{Noronha-Hostler:2008kkf}. We would like to mentioned that the dynamics near the QCD critical point has been previously studied using generalised Langevin equation\cite{Minami:2011un}. In this work, authors have discussed the critical dynamics within the renormalization group approach. Unlike our work, Ref.\cite{Minami:2011un} have taken into account the non-linear couplings between slow modes to calculate the streaming terms. Such couplings play a crucial role in determining the dynamic critical exponents.
    
    We organize the paper as follows. In section \ref{secI} we recapitulate the MZN projection operator method. In section \ref{secII} we present the derivation of density correlation function using MZN projection operator method. In section \ref{secIII} we discuss the density correlation near the QCD critical point within ambit of SBM and  discuss the implications of our results in the context of search of CEP in heavy-ion collision experiments. Finally, in section \ref{secV} we summarize and conclude.

\section{Mori-Zwanzig-Nakajima projection operator method}
\label{secI}
Mori-Zwanzig-Nakajima projection operator method is based on coarse graining a system with large number of microscopic degrees of freedom governed by hamiltonian dynamics into fast degrees of freedom and relatively few effective slow degrees of freedom using time independent projection operator. This procedure leads to so called generalised Langevin equation which  describes the time evolution of a slow operator. While the slow dynamics is captured by so called memory function, the fast dynamics is contained in the noise term.  On the microscopic scale the time dependence of an operator $\mathcal{\hat{O}}$ is governed by Heisenberg equation of motion:

\begin{equation}
\frac{\partial \mathcal{\hat{O}}(t)}{\partial t}=i[\hat{H},\mathcal{\hat{O}}]\equiv i\mathcal{\hat{L}}\mathcal{\hat{O}}(t)
\label{hem}
\end{equation}

where $\mathcal{\hat{L}}\equiv[\hat{H},...]$ is the Liouville operator. $\hat{H}$ is the hamiltonian operator. The formal solution of Eq.(\ref{hem}) is given as,

\begin{equation}
\mathcal{\hat{O}}(t)=e^{it\mathcal{\hat{L}}}\mathcal{\hat{O}}(0),
\end{equation}

where $\mathcal{\hat{O}}(0)$ corresponds to initial time operator. We define inner product of two operators $\mathcal{\hat{A}}$ and $\mathcal{\hat{B}}$ as\cite{mori1965transport,Huang:2011ez},

\begin{eqnarray}
(\mathcal{\hat{A}}, \mathcal{\hat{B}})&=&\frac{1}{\beta}\int_{0}^{\beta} d\tau \: \text{tr}[\rho_0 e^{\tau(\hat{H}-\mu\hat{N})}\mathcal{\hat{A}}e^{-\tau(\hat{H}-\mu\hat{N})}\mathcal{\hat{B}}]\\
&=&\frac{1}{\beta}\int_{0}^{\beta} d\tau <\mathcal{\hat{A}}(-i\tau)\mathcal{\hat{B}}>_0,
\label{KCR}
\end{eqnarray}
where $\beta=T^{-1}$. The equilibrium average is defined as,

\begin{equation}
<\mathcal{\hat{O}}>_0=\text{tr}(\hat{\rho}_0 \mathcal{\hat{O}}).
\end{equation}
 
 For a grand canonical ensemble the density matrix $\rho_0$ is given by,
 
 \begin{equation}
 \hat\rho_0(\beta,\mu)\equiv\frac{e^{-\beta(\hat{H}-\mu\hat{N})}}{\text{tr}\:e^{-\beta(\hat{H}-\mu\hat{N})}}.
 \end{equation}

If it is possible to separate the time scale into long-time and short-time scales then there exists a set of slowly varying operators (corresponding to slow variables) which describes the slow dynamics. Let $\{\hat{\mathcal{A}}_{n}\}=\{\mathcal{\hat{A}}_{1}, \mathcal{\hat{A}}_{2},...,\mathcal{\hat{A}}_{n}\}$  be set of such slowly varying operators which are not necessarily orthogonal. Using Kubo canonical relation given by Eq.(\ref{KCR}) with $\tau=0$  we define metric $g_{nm}$ as\cite{Hidaka:2012ym},
\begin{equation}
\label{rel1}
g_{nm}({\bf{x-y}})\equiv(\mathcal{\hat{A}}_{n}(0,{\bf{x}}),\mathcal{\hat{A}}_{m}(0,{\bf{y}})).
\end{equation}

It is possible now to define  quantity $\mathcal{\hat{A}}^{n}$ orthogonal to $\mathcal{\hat{A}}_{n}$ as,

\begin{equation}
\mathcal{\hat{A}}^n(t,{\bf{x}})=\int d^3 y\:g^{nm}({\bf{x-y}})\mathcal{\hat{A}}_{m},
\end{equation}

where $g^{nm}$ is the inverse of $g_{nm}$ which coincides with the second derivative of the effective action $\beta \Gamma_{\text{eff}}$, with respect to $\mathcal{\hat{A}}_{n}$:

\begin{equation}
g^{nm}({\bf{x-y}})=\frac{\delta^2\beta\Gamma_{\text{eff}}(\mathcal{A}_{n})}{\delta \mathcal{A}_{m}({\bf{y}})\delta\mathcal{A}_{n}({\bf{x}})}
\end{equation}

where the effective action is given by the Legendre transformation of the generating functional $W(J^n)$:

\begin{eqnarray}
\Gamma_{\text{eff}}(\mathcal{A}_{n})&=&W(J^n)-\int d^3x J^{m}({\bf{x}})\frac{\delta W(J^n)}{\delta J^m({\bf{x}})}\\
&=& W(J^n)+\frac{1}{\beta}\int d^3x\: J^{n}({\bf{x}})\mathcal{A}_{n}(0,{\bf{x}})
\end{eqnarray}

where, $\mathcal{A}_{n}({\bf{x}})=<\hat{\mathcal{A}}_{n}({\bf{x}})>_0=\frac{\delta (-\beta W)}{\delta J^n({\bf{x}})}$ and $J^n$ is the classical source.  $W(J^n)=-\frac{1}{\beta}\text{ln}\mathcal{Z}$ with the partition function $\mathcal{Z}$ is given by,

\begin{equation}
\mathcal{Z}(\beta; J^n)=\text{tr}\bigg[e^{-\beta(\hat{H}-\mu\hat{N})}\text{exp}\bigg(\int d^3x \mathcal{\hat{A}}_{n}(0,{\bf{x}})J^n(\bf{x})\bigg)\bigg].
\end{equation}

The quantities with upper index and those with lower indices satisfy following properties:

\begin{equation}
\label{rel2}
(\mathcal{\hat{A}}_{n}(0,{\bf{x}}),\mathcal{\hat{A}}^{m}(0,{\bf{y}})=\delta^{m}_{n}\delta({\bf{x}}-{\bf{y}}),
\end{equation}

 \begin{equation}
 \label{rel3}
 \sum_{p}\int d^3z\:g_{mp}(x-z)g^{pn}(z-y)=\delta^{n}_{m}\delta({\bf{x}}-{\bf{y}}).
 \end{equation}

We finally define the projection operator $\mathcal{\hat{P}}$ acting on any arbitrary operator $\mathcal{\hat{O}}(t,{\bf{x}})$ as,

\begin{equation}
\label{projdef}
\mathcal{\hat{P}}\mathcal{\hat{O}}(t,{\bf{x}})\equiv\int d^3z\: \mathcal{\hat{A}}_{n}(0,{\bf{z}})(\mathcal{\hat{O}}(t,{\bf{x}}),\mathcal{\hat{A}}^{n}(0,{\bf{z}})).
\end{equation}

Operator $\mathcal{\hat{P}}$ project out the slowly varying part of $\mathcal{\hat{O}}$. Using (\ref{rel1}), (\ref{rel2}) and (\ref{rel3}) it can be easily shown that (\ref{projdef}) satisfies $\mathcal{\hat{P}}^2=\mathcal{\hat{P}}$. It is also useful to define orthogonal projector $\mathcal{\hat{Q}}\equiv 1-\mathcal{\hat{P}}$. This project out the part of Hilbert space orthogonal to subspace occupied by slowly varying operators.

Now consider an operator identity:

\begin{equation}
\frac{\partial}{\partial t}e^{it\mathcal{\hat{L}}}=e^{it\mathcal{\hat{L}}}\mathcal{\hat{P}}\:i\mathcal{\hat{L}}+\int_{0}^{t} ds\:e^{i(t-s)\mathcal{\hat{L}}}\mathcal{\hat{P}}\:i\mathcal{\hat{L}}\:e^{it\mathcal{\hat{Q}}\mathcal{\hat{L}}}\:\mathcal{\hat{Q}}\:i\mathcal{\hat{L}}+e^{it\mathcal{\hat{Q}}\mathcal{\hat{L}}}\:\mathcal{\hat{Q}}\:i\mathcal{\hat{L}}
\label{opid}
\end{equation}

Multiplying both sides of (\ref{opid}) by $\mathcal{\hat{A}}_n(0)$ we get

\begin{eqnarray}
\label{langevin}
\frac{\partial}{\partial t}\mathcal{\hat{A}}_n(t,{\bf{x}})&=&\int d^3y\: iK_n^{(s)m}({\bf{x-y}})\mathcal{\hat{A}}_{m}(t,{\bf{y}})\\\nonumber
&-&\int_{0}^{\infty}ds\int d^3y\: K_n^{(d)m}(t-s,{\bf{x-y}})\mathcal{\hat{A}}_{m}(s,{\bf{y}})+\mathcal{\hat{N}}_{n}(t,{\bf{x}}),
\end{eqnarray}

where,

\begin{eqnarray}
\label{smf}
iK_n^{(s)m}({\bf{x-y}})&\equiv& (i\mathcal{\hat{L}}\mathcal{\hat{A}}_{n}(0,{\bf{x}}), \mathcal{\hat{A}}^{m}(0,{\bf{y}}))\\
\label{dmf}
K_n^{(d)m}(t-s,{\bf{x-y}})&\equiv&-\theta(t-s) (i\mathcal{\hat{L}}\mathcal{\hat{N}}_{n}(t,{\bf{x}}), \mathcal{\hat{A}}^{m}(s,{\bf{y}}))\\
\label{noise}
\mathcal{\hat{N}}_n(t,{\bf{x}})&\equiv& e^{it\mathcal{\hat{Q}}\mathcal{\hat{L}}}\:\mathcal{\hat{Q}}\:i\mathcal{\hat{L}}\mathcal{\hat{A}}_{n}(0,{\bf{x}}).
\end{eqnarray}

Eq.(\ref{langevin}) is called generalized Langevin equation\cite{mori1965transport}. $iK_n^{(s)m}$ is called streaming term and it captures the time-reversible change. While, $K_n^{(d)m}$ is called dynamic memory function and it captures the time-irreversible change and hence the dissipation in the system. Note that  $K_n^{(d)m}$ depends on a past time value of $\mathcal{\hat{A}}_{n}(s)$ for $s<t$. The last term in the generalized Langevin equation $\mathcal{\hat{N}}$ is called the noise. It is only the memory function terms that contribute to the slow dynamics and hence we neglect the noise term hereafter.

Taking Fourier transform of (\ref{langevin}) we get,

\begin{eqnarray}
\label{langevink}
\frac{\partial}{\partial t}\mathcal{\hat{A}}_n(t,{\bf{k}})&=&\: iK_n^{(s)m}({\bf{k}})\mathcal{\hat{A}}_{m}(t,{\bf{k}})\\\nonumber
&-&\int_{0}^{\infty}ds\: K_n^{(d)m}(t-s,{\bf{k}})\mathcal{\hat{A}}_{m}(s,{\bf{k}})+\mathcal{\hat{N}}_{n}(t,{\bf{k}}).
\end{eqnarray}

Eq.(\ref{langevink}) can we written in more compact form as,

\begin{equation}
\frac{\partial}{\partial t}\mathcal{\hat{A}}(t)=\: i{\bf{K^{(s)}}}\mathcal{\hat{A}}(t)-\int_{0}^{\infty}ds\: {\bf{K^{(d)}}}(t-s){\bf{\mathcal{\hat{A}}}}(s)+\mathcal{\hat{N}}(t)
\label{langevink1}
\end{equation}

In the Eq.(\ref{langevink1}) $\mathcal{\hat{A}}$ is a column matrix of order $n$ for a set of '$n$' slow variables. Matrices $\bf{K^{(s)}}$ and $\bf{K^{(d)}}$ are $n\times n$ matrices.

\section{Density correlations for the equilibrium fluctuations }
\label{secII}
We shall now calculate the density correlations for the fluctuations about the thermal equilibrium. Near the thermal equilibrium any thermodynamic quantity at an arbitrary spacetime point $(t, {\bf{x}})$ can be written as,

\begin{equation}
A(t,{\bf{x}})=A_0+\delta A(t,{\bf{x}}),
\end{equation}
 where $A_0$ is the equilibrium value and $\delta A(t,{\bf{x}})$ is the fluctuation about equilibrium. For a baryon-rich fluid the relevant slow variables are fluctuations in baryon number density $\delta\hat{n}_{b}$, energy density $\delta\hat{\varepsilon}$, pressure $\delta \hat{P}$, entropy density $\delta \hat{s}$ and fluid velocity $\delta \hat{\bf{v}}$. However, as we will see, not all of these are independent and we can remove two of them. So we are left with five independent slow variables. For our purpose we choose  $\delta\hat{n}_b$,  $\delta\hat{\epsilon}$ and the fluid velocity $\delta\hat{{\bf{v}}}$ as slow variables. As we will see this choice of slow variables leads to Landau form of hydrodynamic equations. This also implies that the fluid velocity is defined with respect to Landau frame.

 The generalised Langevin equation for $\delta \hat n_{b}$ reads:

\begin{eqnarray}
\label{langevin1}
\frac{\partial}{\partial t}\delta \hat{n}_b(t,{\bf{x}})&=&\int d^3y\:iK_{{n}_b}^{(s)m}({\bf{x-y}})\mathcal{\hat{A}}_{m}(t,{\bf{x}})\\\nonumber
&-&\int_{0}^{\infty}ds\int d^3y\: K_{n_b}^{(d)m}(t-s,{\bf{x-y}})\mathcal{\hat{A}}_{m}(s,{\bf{x}}).
\end{eqnarray}

 Using memory matrices calculated in the Appendix \ref{appendix1} we get,
 
 \begin{equation}
\frac{\partial \delta \hat{n}_{b}}{\partial t}+\hat{n}_{b,0}\nabla\cdot\delta{\bf{\hat{v}}}-\kappa \bigg(\frac{\hat{n}_{b,0}T_0}{h_0}\bigg)^2\nabla^2\delta(\hat{\beta}\hat{\mu}_b)=0,
\label{landau1}
\end{equation}
where, $h_0$ is the equilibrium enthalpy density. A similar calculation for $\delta \hat{{\bf{v}}}$ and  $\delta \hat{\epsilon}$  respectively gives,

\begin{equation}
h_0\frac{\partial \delta {\bf{\hat{v}}}}{\partial t}+\nabla\delta \hat{P}-\eta\nabla^2\delta\hat{\bf{v}}-(\zeta+\frac{1}{3}\eta)\nabla(\nabla\cdot\delta{\bf{\hat{v}}})=0,
\label{landau2}
\end{equation}

\begin{equation}
\frac{\partial \delta {\hat{\epsilon}}}{\partial t}+h_0\nabla\cdot\delta\hat{{\bf{v}}}=0.
\label{landau3}
\end{equation}

 Using following thermodynamic relations:

\begin{eqnarray}
\delta\epsilon&=&T_0\delta(n_b s)+\mu_{b,0}\delta n_b,\\
\delta P&=&n_{b,0}s_0\delta T+n_{b,0}\delta\mu_B,\\
h_0&=&T_0n_{b,0}s_0+n_{b,0}\mu_{b,0},
\end{eqnarray}

and choosing $\delta \hat{n}_{b}$ and $\delta \hat{T}$ as independent variables, we can  rewrite Eqs. (\ref{landau1})-(\ref{landau3}) in the form:

\begin{eqnarray}
\label{landau4}
\bigg[\frac{\partial }{\partial t}-\kappa\bigg(\frac{T_0C_s^2\tilde{C}_n}{h_0\tilde{C}_P}\bigg)\nabla^2\bigg]\delta\hat{n_b}+\kappa\bigg[1-\frac{\tilde{C}_nC_s^2\alpha_PT_0}{\tilde{C}_P}\bigg]\nabla^2\delta\hat{T}+n_{b,0}\nabla\cdot\delta\hat{\bf{v}}&=&0,\\
\label{landau5}
\bigg(\frac{h_0\tilde{C_n}C_s^2}{n_{b,0}\tilde{C}_P}\bigg)\nabla\delta\hat{n}_b+\bigg(\frac{h_0\tilde{C_n}C_s^2\alpha_P}{\tilde{C}_P}\bigg)\nabla\delta\hat{T}+\bigg(h_0\frac{\partial}{\partial t}-\eta\nabla^2\bigg)\delta\hat{{\bf{v}}}-\bigg(\zeta+\frac{1}{3}\eta\bigg)\nabla(\nabla\cdot\delta\hat{\bf{v}})&=&0,\\
\label{landau6}
\bigg[\bigg(\frac{h_0C_s^2\alpha_P\tilde{C}_n}{n_{b,0}\tilde{C}_P}\bigg)\frac{\partial }{\partial t}-\kappa\bigg(\frac{\tilde{C}_nC_s^2\tilde{C}_n}{n_{b,0}\tilde{C}_P}\bigg)\nabla^2\bigg]\delta\hat{n}_b-\bigg[\frac{n_{b,0}\tilde{C}_n}{T_0}\frac{\partial }{\partial t}+\kappa\bigg(\frac{\alpha_PC_s^2\tilde{C}_n}{\tilde{C}_P}-\frac{1}{T_0}\bigg)\nabla^2\bigg]\delta\hat{T}&=&0,
\end{eqnarray}

where we have used following relations:

\begin{eqnarray}
\delta P&=&\bigg(\frac{\partial P}{\partial n_b}\bigg)_T\delta n_b+\bigg(\frac{\partial P}{\partial T}\bigg)_{n_{b}}\delta T,\\\nonumber
\delta s&=&\bigg(\frac{\partial s}{\partial n_b}\bigg)_T\delta n_b+\bigg(\frac{\partial s}{\partial T}\bigg)_{n_{b}}\delta T,
\label{eulerder}
\end{eqnarray} 

 together with the thermodynamic identities\cite{Minami:2009hn}:
 \begin{eqnarray}
 \bigg(\frac{\partial P}{\partial n_b}\bigg)_T&=&\frac{h_0C_s^2\tilde{C}_{n_b}}{\tilde{C}_Pn_{b,0}},\\
 \bigg(\frac{\partial P}{\partial T}\bigg)_{n_{b}}&=&\frac{h_0\alpha_P C_s^2\tilde{C}_n}{\tilde{C}_P},\\
 \bigg(\frac{\partial s}{\partial n_b}\bigg)_T&=&-\frac{h_0\alpha_PC_s^2\tilde{C}_{n_b}}{n_{b,0}^2\tilde{C}_P},\\
 \bigg(\frac{\partial s}{\partial T}\bigg)_{n_{b}}&=&\frac{\tilde{C}_n}{T_0},
 \end{eqnarray}

where, $C_s^2=(\frac{\partial P}{\partial \epsilon})_s$ is the speed of sound, $\tilde{C}_{n_b}=T_0(\frac{\partial s}{\partial T})_{n_b}$ is the specific heat at constant number density, $\tilde{C}_{P}=T_0(\frac{\partial s}{\partial T})_P$ is the specific heat at constant pressure, $\alpha_P=-\frac{1}{n_{b,0}}(\frac{\partial n_b}{\partial T})_P$ is the thermal expansivity at constant pressure.

It is convenient to transform Eqs.(\ref{landau4})-(\ref{landau6}) using Laplace-Fourier transform:

\begin{equation}
\tilde{f}(z,{\bf{k}})=\int_{-\infty}^{+\infty} d{\bf{x}}\:e^{i{\bf{k\cdot x}}}\int_{0}^{\infty} dt\: e^{-zt}\: f(t,{\bf{x}}).
\end{equation}

Eqs.(\ref{landau4})-(\ref{landau6}) can now be written in a matrix form:

\begin{gather}
X
\begin{bmatrix} 
	\delta\hat{n}_b(z, {\bf{k}})  \\
	\delta\hat{T}(z, {\bf{k}})  \\
	\delta\hat{{\bf{v_{\parallel}}}}(z, {\bf{k}})  \\
	\delta\hat{{\bf{v_{\perp}}}}(z, {\bf{k}}) \\
	\end{bmatrix}
	=
	\begin{bmatrix} 
	\delta\hat{n}_b(0, {\bf{k}})  \\
	\frac{n_{b,0}\tilde{C}_n}{T_0h_0}\delta\hat{T}(0, {\bf{k}})-\frac{\alpha_P\tilde{C}_nC_S^2}{n_{b,0}\tilde{C}_{P}}\delta\hat{n}_b(0, {\bf{k}})   \\
	\delta\hat{{\bf{v_{\parallel}}}}(0, {\bf{k}})  \\
	\frac{h_0}{zh_0+k^2\eta}\delta\hat{{\bf{v_{\perp}}}}(0, {\bf{k}}) \\
	\end{bmatrix}
	\label{matrixeqn1}
\end{gather}	
 or
 
 \begin{equation}
 X\hat{A}_{n}(z, {\bf{k}})=\hat{A}_{n}(t=0, {\bf{k}}),
 \label{matrixeqn2}
 \end{equation}

where,

\begin{gather}
X=\begin{bmatrix} 
	z+\kappa\frac{T_0C_s^2\tilde{C}_n}{h_0\tilde{C}_P}k^2 & \bigg[-\kappa\frac{n_{b,0}}{h_0}(1-\frac{\alpha_P\tilde{C}_nC_S^2T_0}{\tilde{C}_P})k^2\bigg] & in_{b,0}k & 0 \\
	i\frac{\tilde{C_n}C_s^2}{n_{b,0}\tilde{C}_P}k & i\frac{\tilde{C_n}C_s^2\alpha_P}{\tilde{C}_P}k & z+\Omega_lk^2 & 0\\
	\bigg[-\bigg(\frac{h_0C_s^2\alpha_P\tilde{C}_n}{n_{b,0}\tilde{C}_P}\bigg)z-\kappa\bigg(\frac{\tilde{C}_nC_s^2\tilde{C}_n}{n_{b,0}\tilde{C}_P}\bigg)k^2\bigg] &\frac{n_{b,0}\tilde{C}_n}{h_0T_0}\bigg[z+\frac{h_0\tilde{C}_P}{n_{b,0}\tilde{C}_n} D_T\bigg(1-\frac{T_0\alpha_PC_s^2\tilde{C}_n}{\tilde{C}_P}\bigg)k^2\bigg] & 0 & 0 \\
	0 & 0 & 0 & 1\\
	\end{bmatrix}
\end{gather}	

where we have decomposed the fluid velocity parallel ($\hat{\bf{v}}_{\parallel}$) and  perpendicular ($\hat{\bf{v}}_{\perp}$) to the wave vector ${\bf{k}}$. The quantities $D_T$ and $\Omega_l$ are given by,

\begin{eqnarray}
D_T&=&\frac{\kappa}{h_{0}\tilde{C}_P},\\
\Omega_l&=&\frac{1}{h_0}\bigg(\zeta+\frac{4}{3}\eta\bigg).
\end{eqnarray}

 Quantities $D_T$ and $\Omega_l$ are, respectively, called thermal diffusivity and longitudinal viscosity. Note that the transverse velocity fluctuation is decoupled from the density fluctuation. So we shall ignore this mode in present analysis. We Replace 4-dimensional matrix equation (\ref{matrixeqn2}) by 3-dimensional matrix equation as,
 
  \begin{equation}
 \mathcal{X}\mathcal{\hat{A}}_{n}(z, {\bf{k}})=\mathcal{\hat{A}}_{n}(t=0, {\bf{k}}),
 \label{matrixeqn3}
 \end{equation}
which can be obtained from  (\ref{matrixeqn1}) by deleting bottom row.

 Dynamical density correlation function is defined as,
 
 \begin{equation}
 \mathcal{C}_{n_{b}n_{b}}(\omega,{\bf{k}})\equiv <\delta\hat{n}_b(\omega,{\bf{k}})\delta\hat{n}_b(t=0,{\bf{k}})>,
 \end{equation}

where $\delta\hat{n}_b(\omega,{\bf{k}})$ is the Fourier transform of $\delta\hat{n}_b(t,{\bf{k}})$ whereas later can be obtained by taking inverse Laplace's transform of $ \delta\hat{n}_b(z,{\bf{k}})$ as,

\begin{equation}
\delta\hat{n}_b(t,{\bf{k}})=\frac{1}{2\pi i}\int_{\delta-i\infty}^{\delta-i\infty}dz\: e^{zt} \delta\hat{n}_b(z,{\bf{k}}).
\end{equation}

We first calculate $\delta\hat{n}_b(z,{\bf{k}})$ by inverting matrix equation (\ref{matrixeqn3}) as,

 \begin{equation}
 \mathcal{\hat{A}}_{n}(z, {\bf{k}})=\mathcal{X}^{-1}\mathcal{\hat{A}}_{n}(t=0, {\bf{k}}).
 \label{matrixeqn4}
 \end{equation}

The Fourier-Laplace space density function $\delta\hat{n}_b(z,{\bf{k}})$ is the (1,1) component of $\hat{A}_{n}$.

\begin{equation}
\delta\hat{n}_b(z,{\bf{k}})=( \mathcal{\hat{A}}_{n}(z, {\bf{k}}))_{11}=\sum_{j=1}^{4}\mathcal{X}^{-1}_{1j}\mathcal{\hat{A}}_{n}(t=0, {\bf{k}})_{j1}.
\label{A11}
\end{equation}

The inverse of $\mathcal{X}$ can be easily obtained one we know its determinant. At the leading order the determinant of $\mathcal{X}$ is,

\begin{equation}
\text{det}\mathcal{X}=\frac{n_{b,0}\tilde{C}_n}{h_0T_0}\bigg\{z^3+z^2k^2\bigg[\bigg(\frac{h_0\tilde{C}_P}{n_0\tilde{C}_n}+\tilde{C}_PC_s^2T_0-2C_s^2\alpha_PT_0\bigg)D_T+\Omega_l\bigg]+zk^2C_s^2+\mathcal{O}(k^4)\bigg\}.
\label{det1}
\end{equation}

The Fourier-Laplace coefficient  of density correlation (\ref{A11}) at the order $\mathcal{O}(k^2)$ becomes,

\begin{equation}
\label{deltanb}
\frac{\delta \hat{n}_{b}(z,{\bf{k}})}{\delta \hat{n}_{b}(0,{\bf{k}})}\simeq \frac{(z+\Gamma_Sk^2+iC_sk)(z+\Gamma_Sk^2-iC_sk)+zk^2D_T\tilde{C}_nC_s^2-k^2C_s^2\frac{\tilde{C}_n}{\tilde{C}_P}}{(z+\Gamma_Sk^2+iC_sk)(z+\Gamma_Sk^2-iC_sk)(z+\frac{h_0}{n_{b,0}}D_Tk^2)},
\end{equation}

where,

\begin{eqnarray}
\label{ray}
\Gamma_{T}&=&\frac{h_0}{n_{b,0}}D_T,\\
\label{bril}
\Gamma_{S}&=&\frac{1}{2}\bigg\{\Omega_l+D_T\bigg[\bigg(\frac{\tilde{C}_P}{\tilde{C}_n}-1\bigg)+C_s^2T_0(\tilde{C}_P-2\alpha_P)\bigg]\bigg\}.
\end{eqnarray}
In Eq.(\ref{deltanb}) we have neglected the dependence of $\delta\hat{n}_{b}$ on $\delta \hat{\bf{v}}$ and $\delta\hat{\varepsilon}$ because the definition of density correlation function innvolve the thermal averaging and averages,  $<\delta \hat{n}_{b}(k,0)\delta \hat{\varepsilon} (k,0)>$ and $<\delta n_{b}(k,0)\delta \hat{\bf{v}}(k,0)>$ vanish\cite{reichl1998modern}. Taking inverse Fourier-Laplace transform and then averaging over thermal equilibrium we finally obtain the dynamical density correlation function as,

\begin{eqnarray}
\label{corfun}
\mathcal{C}_{n_bn_b}(\omega, {\bf{k}})&=&<(\delta \hat{n}_{b}(t=0,{\bf{k}}))^2>\bigg[\bigg(1-\frac{\tilde{C}_n}{\tilde{C}_P}\bigg)\frac{2\Gamma_T k^2}{\omega^2+\Gamma_T^2 k^4}\\\nonumber
&+&\frac{\tilde{C}_n}{\tilde{C}_P}\bigg(\frac{\Gamma_S k^2}{(\omega-C_s k)^2+\Gamma^2 k^4}+\frac{\Gamma_S k^2}{(\omega+C_s k)^2+\Gamma^2_S k^4}\bigg)\bigg].
\end{eqnarray}

One can derive similar form using linearised hydrodynamic equations in the Landau frame\cite{Minami:2009hn}. The density correlation function has three peaks as in the non-relativistic case\cite{mazenko2008nonequilibrium}. The peak around $\omega=0$ is called Rayleigh peak which corresponds to thermally induced density fluctuations. We note that the width of Rayleigh peak $\Gamma_T$ (Eq.(\ref{ray})) is the same as that of non-relativistic case except for the pre-factor $\frac{h_0}{n_{b,0}}$. The two peaks at $\omega=\pm C_sk$ corresponds to the acoustic waves and they are called Brillioun peaks. Note that there are additional terms (see the discussion below) in the width of Brillioun peaks $\Gamma_S$ (Eq.(\ref{bril})) which are purely relativistic\cite{Minami:2009hn}. These terms are proportional to the thermal diffusivity. The origin of additional terms appearing in the relativistic expression for density correlation function  (\ref{corfun}) as compared to the non-relativistic one lies in the hydrodynamic equations (\ref{landau1}), (\ref{landau2}) and (\ref{landau3}). Being relativistic generalisation of non-relativistic hydrodynamic equations,  the mass density appearing in the non-relativistic hydrodynamic equations is replaced enthalpy density $h_0$ and there are additional pressure gradient terms which are absent in the non-relativistic case.

To bring out the genuine relativistic effects in the density correlations consider a  gas of massless particles. In the massless limit $\zeta=0$ . Thus only shear viscosity and thermal conductivity terms contribute to the sound attenuation as,

\begin{equation}
\Gamma_S=\frac{1}{2}\bigg[\frac{h_0}{n_0}D_T\bigg(\frac{\tilde{C}_{P}}{\tilde{C}_{n}}-1\bigg)+\frac{4}{3}\frac{\eta}{h_0}+C_s^2 T_0\bigg(\frac{\kappa}{h_0}-\frac{2\alpha_Ph_0}{n_0} D_T\bigg)\bigg]=\Gamma_{S,R}+\delta\Gamma_{\text{S,R}},
\end{equation}

where,

\begin{eqnarray}
\Gamma_{S,R}&=&\frac{1}{2}\bigg[\frac{h_0}{n_0}D_T\bigg(\frac{\tilde{C}_{P}}{\tilde{C}_{n}}-1\bigg)+\frac{4}{3}\frac{\eta}{h_0}\bigg],\\
\delta\Gamma_{\text{S,R}}&=&C_s^2 T_0\bigg(\frac{\kappa}{h_0}-\frac{2\alpha_Ph_0}{n_0} D_T\bigg),
\end{eqnarray}

 where $\Gamma_{S,R}$ is the relativistic counterpart of sound attenuation.  $\delta\Gamma_{\text{S,R}}$ is a purely relativistic correction.  In the extreme relativistic limit, i.e when $T_0$ is very large or when the kinetic energy dominates the rest-mass energy, we can approximate the gas by its ideal limit. For a massless ideal classical gas, we have $\tilde{C}_P=4, \tilde{C}_n=3, \alpha_P=1/T_0, C_s^2=1/3$.  In this limit, $\delta\Gamma_{\text{S,R}}\simeq -\frac{\kappa}{24n_{b,0}}$. On the other hand,  $\delta\Gamma_{\text{S,R}} \rightarrow 0$ in the non-relativistic limit ($T_0\rightarrow 0$ or when the rest-mass energy dominates the kinetic energy). Thus the relativistic correction to the sound attenuation is always negative. This implies that for a relativistic fluid the Brillouin peaks are enhanced and its width reduced by an amount $\delta\Gamma_{\text{R}}$ as compared to non-relativistic counterpart. On the other hand,  the width Rayleigh peak is practically remain unaltered for the relativistic systems. Reason for this is twofold. First, there are no relativistic corrections to $\Gamma_T$. Second, albeit the thermal diffusivity of the relativistic system ($D_T=\frac{\kappa}{h_{0}\tilde{C}_P}$) is different from its non-relativistic counterpart ($D^{NR}_T=\frac{\kappa}{n_{b,0}\tilde{C}_P}$), the expression for Rayleigh peak, $\Gamma_T=\frac{\kappa}{n_{b,0}\tilde{C}_P}=D_T$ exactly matches its non-relativistic counterpart. Hence the the strength and the width of the Rayleigh peak is not affected by the relativistic effects.
 
 Finally, for a massless ideal classical gas case, the width of Brillouin peak reduces to $ \Gamma_{S,R}=\frac{\eta}{6n_{b,0}T_0}$. Thus, the bulk viscosity and the thermal conductivity  does not have a net effect on the Brillouin peaks. This is because of the cancellation of terms involving thermal conductivity and because $\zeta=0$ for the massless gas.

\section{Density correlations near the critical point}
\label{secIII}

In the $T \rightarrow T_c$ limit, any relevant thermodynamic quantity can be separated into a regular part and a singular part. The singular part show power law behaviour. This behaviour is universal and characterized by   the critical exponents, $\hat{\alpha}$, $\hat{\beta}$, $\hat{\gamma}$ and $\hat{\nu}$. They are defined through the following power laws~\cite{Huang:1987sm} (in the limit $t \rightarrow 0^{-}$):
\begin{eqnarray}
\label{alpha_exponent}
C_V &=&\mathcal{C_{-}} \: |t|^{-\hat{\alpha}}, \\
\label{beta_exponent}
\tilde{n}_{b}&=& \mathcal{N_-} \: |t|^{\hat{\beta}}, \\ 
\label{gamma_exponent}
k_{T} &=&\mathcal{K}_{-} \: |t|^{-\hat{\gamma}}, \\
\label{nu_exponent}
\xi &=&\Xi_{-}\:|t|^{-\hat{\nu}},
\end{eqnarray}
where, $\tilde{n}_{b}=1-\frac{n_{b}}{n_{b,c}}$ is the order parameter with $n_{b,c}$ being critical density. $C_V$, $k_{T}$ and $\xi$ respectively denote the specific heat, the isothermal compressibility and the correlation length. $\mathcal{C_{-}}$, $\mathcal{N_-}$, $\mathcal{K}_{-}$ and $\Xi_{-}$ are the corresponding amplitudes from the hadronic side ($T <T_c$). Note that the correlation length $\xi$ in Eq.(\ref{nu_exponent}) is the typical length scale of hadronic interactions. So away from the critical point $\Xi_{-}\sim 1$ fm.

In the Statistical Bootstrap Model (SBM)  the strong interactions are assumed to be simulated by the presence of hadronic clusters\cite{Hagedorn:1965st,Frautschi:1971ij}. The hadronic states of this cluster are described by exponentially rising density of states.  The thermodynamic properties of SBM can be derived using following partition function\cite{Kadam:2020utt}:

\begin{equation}
\text{ln}\mathcal{Z}_H(T,V,z_b)=AVz_b\left(\frac{T}{2\pi}\right)^{3/2}\int_{M}^{\infty}dm\:m^{a+3/2}  e^{\big(\frac{1}{T_c}-\frac{1}{T}\big) m}.
\label{pf4}
\end{equation}
where, $z_{b}=e^{\mu_{b}/T}$ is the fugacity. The exponent plays a very important role in determining the thermodynamic behaviour of hadronic matter near the critical point.  It can be shown that for $a<-\frac{7}{2}$ SBM shows critical behaviour\cite{Satz:1978us}. It turns out that the energy density and entropy remains finite as $T\rightarrow T_c$ whereas all the higher order derivatives show singular behaviour. In fact, it is straightforward to extract the power law behaviour given by Eqns (\ref{alpha_exponent})-(\ref{gamma_exponent}).

Approaching the critical point the functional dependence of transport coefficients on various thermodynamic quantities can be obtained by dimensional analysis.  We assume following ansatz (in the natural units) for the shear viscosity ($\eta$), bulk viscosity ($\zeta$)\cite{Antoniou:2016ikh} and thermal conductivity ($\kappa$):

\begin{eqnarray}
\label{ansatz}
\eta &=&\frac{T}{\xi^2 C_{s}}\:\mathcal{F}^{(\eta)}\bigg(\frac{\tilde{C}_p}{\tilde{C}_n}\bigg), \\
\zeta &=& h \xi C_s\:\mathcal{F}^{(\zeta)}\bigg(\frac{\tilde{C}_p}{\tilde{C}_n}\bigg), \\
\kappa &=& \frac{h \xi T^2  C_s}{\tilde{n}_{B}}\:\mathcal{F}^{(\kappa)}\bigg(\frac{\tilde{C}_p}{\tilde{C}_n}\bigg).
\end{eqnarray}

Substituting power law behaviour of the thermodynamic quantities given by Eqns.(\ref{alpha_exponent})-(\ref{nu_exponent}) in Eq.(\ref{ansatz}) one can obtain the singular behaviour of the transport coefficients. The leading term is

\begin{eqnarray}
\label{etatc}
\eta &\sim & |t|^{-\hat{\gamma}+2\hat{\nu}+\hat{\alpha}/2},\\
\label{zetatc}
\zeta &\sim & |t|^{-\hat{\gamma}-\hat{\nu}+3\hat{\alpha}/2},\\
\label{kappatc}
\kappa &\sim & |t|^{-\hat{\beta}-\hat{\gamma}-\hat{\nu}+3\hat{\alpha}/2}.
\end{eqnarray} 

Note that not all the critical exponents are independent but are related via. scaling laws:

\begin{eqnarray}
\label{scaling}
2-\hat{\alpha}&=&\hat{\nu}d \\
\hat{\alpha}+2\hat{\beta}+\hat{\gamma}&=&2,
\end{eqnarray}
where $d$ is the number of space dimensions. For the choice $a=-4$\cite{Kadam:2020utt,Antoniou:2002xq} we get the critical exponents of SBM as:

\begin{equation}
\hat{\alpha}=\frac{1}{2}, \hspace{0.2cm} \hat{\beta}=0, \hspace{0.2cm} \hat{\gamma}=\frac{3}{2},  \hspace{0.2cm}  \hat{\nu}=\frac{1}{2}.
\label{ce}
\end{equation}

 Substituting these critical exponents  in (\ref{etatc})-(\ref{kappatc}), we get

\begin{eqnarray}
\label{visce}
\eta &=& \eta_-\: |t|^{-\frac{1}{4}},\\\nonumber
\zeta &= & \zeta_-\: |t|^{-\frac{5}{4}},\\\nonumber
\kappa &= & \kappa_-\: |t|^{-\frac{5}{4}},
\end{eqnarray}
where $\eta_-,\zeta_-, \kappa_-$ are constants. Thus the bulk viscosity and thermal conductivity dominates the dissipation near the critical point. The thermal diffusivity behaves near the critical point as

\begin{equation}
D_{T}\sim\frac{\kappa_-}{\mathcal{K}_-T_0}|t|^{\hat{\beta}-\hat{\nu}+3\hat{\alpha}/2}.
\end{equation}

Thus,

\begin{equation}
D_{T}\sim |t|^\frac{1}{4},
\end{equation}

where we have substituted the critical exponents (\ref{ce}) of SBM mode. This shows that the width of Rayleigh peak becomes narrow as the QCD critical point is approached from the hadronic side. Similarly, the the leading order behaviour of the sound attenuation coefficient is:

\begin{equation}
\Gamma\sim K_1\kappa_{-}|t|^{-\hat{\beta}-\hat{\gamma}-\hat{\nu}+\frac{5}{2}\hat{\alpha}}+K_2\zeta_{-}|t|^{-\hat{\gamma}-\hat{\nu}+\frac{3}{2}\hat{\alpha}}+K_3\eta_{-}|t|^{-\hat{\gamma}+2\hat{\nu}+\frac{\hat{\alpha}}{2}},
\end{equation} 
 where $K_1, K_2, K_3$ contains quantities which are not singular.  Substituting the critical exponents of SBM we see that the leading order behaviour of the singularity near the critical point is dominated by the bulk viscosity while the thermal conductivity contributes at sub-leading order.
 
 \begin{equation}
 \Gamma\sim \zeta_-|t|^{-\frac{5}{4}}+\kappa_-|t|^{-\frac{3}{4}}.
 \end{equation}

Thus the width of Brillouin peaks must diverge near at the critical point. However, the strengths of the Rayleigh and  Brillouin peaks is governed by the ratio of two specific heats $\gamma=\frac{\tilde{C}_P}{\tilde{C}_{n}}$ which behave in the limit $T\rightarrow T_c$ as

\begin{equation}
\gamma\sim |t|^{-\hat{\gamma}+\hat{\alpha}}.
\end{equation}

 In the limit $T\rightarrow T_c$, $\gamma\rightarrow\infty$ and  the density correlation behaves as,

\begin{equation}
\frac{\mathcal{C}_{n_bn_b}(\bf{k}, \omega)}{<(\delta n({\bf{k}},t=0))^2>}=\frac{2\Gamma_{T}k^2}{\omega^2+\Gamma_{T}^2k^4},].
\end{equation}

 Thus the strength of the Brillouin peaks is attenuated and only the Rayleigh peak governs the density correlations near the critical point.

\begin{figure}[h!]
\begin{center}
\centerline{\includegraphics[width=0.5\textwidth]{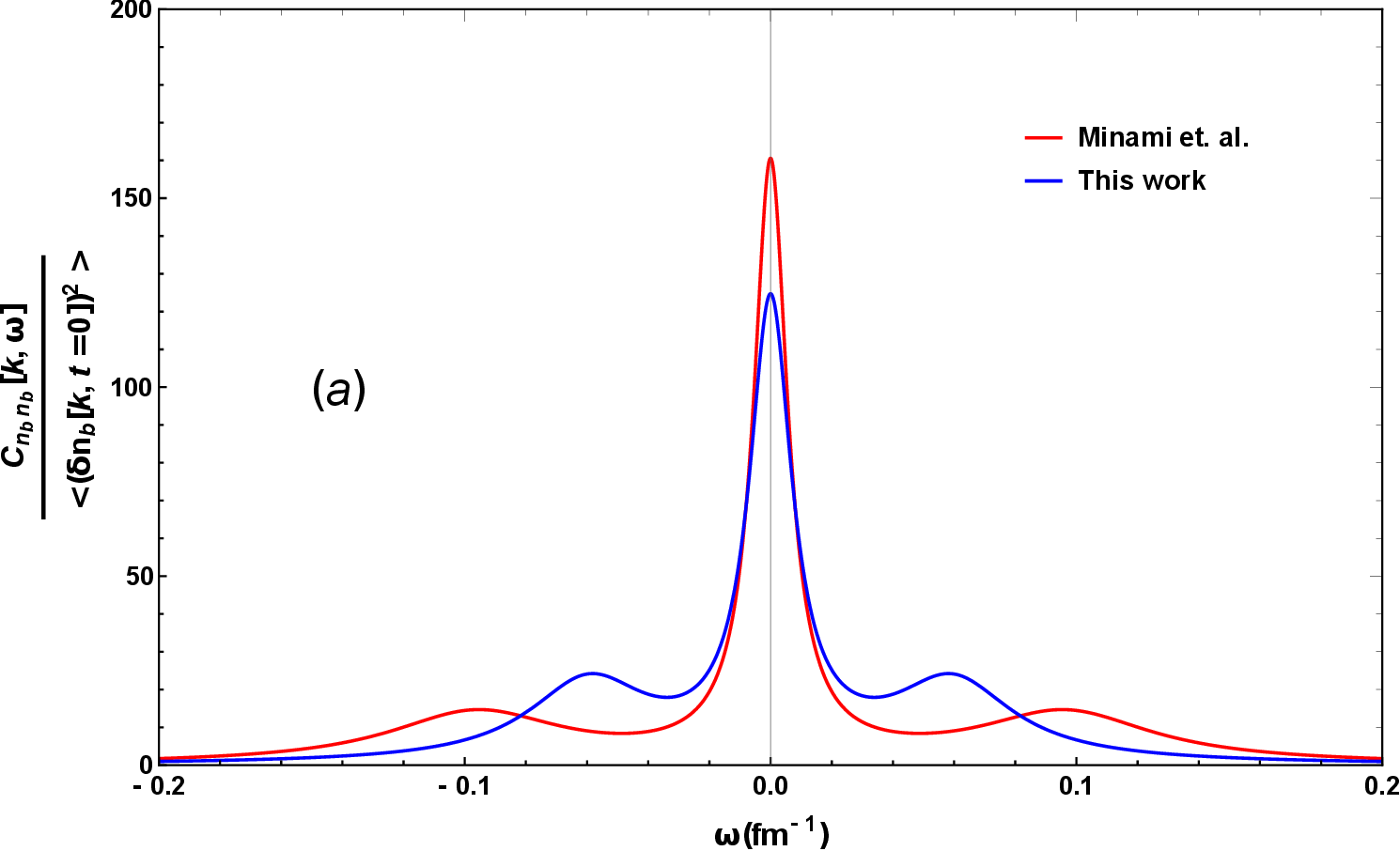} \hfill \includegraphics[width=0.5\textwidth]{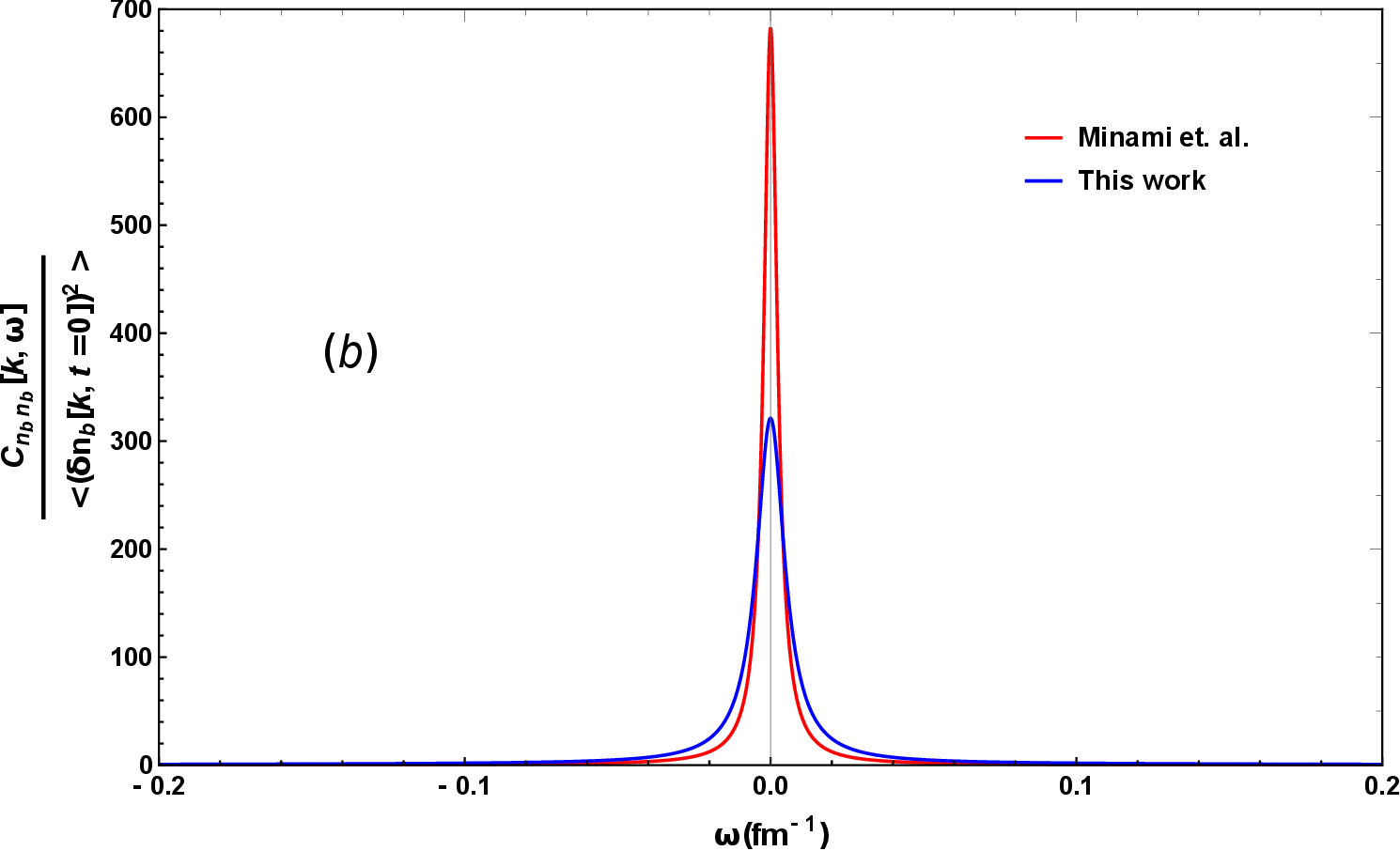}}
		\caption{Correlation function near the QCD critical point for $k=0.1$ fm$^{-1}$. We choose two representative vales of $t=\frac{T-T_c}{T_c}$, namely $t=0.5$ (left panel) and $t=0.1$ (right panel). Red curve represents the results of Ref.\cite{Minami:2009hn} whereas the blue curve represents the results of this work.  }
		\label{cnn}
	\end{center}
\end{figure}

  Fig. \ref{cnn} shows the density correlation function near the QCD critical point. Red curve represents the results of \cite{Minami:2009hn} in which the correlation function has been estimated based on the critical exponents of 3-d Ising model and the singular behaviour of the transport coefficients based on arguments of Ref.\cite{onuki1997dynamic}. The blue curve represents the results corresponding to the critical exponents of SBM and the singular behaviour of the transport coefficients given by Eqs.(\ref{etatc})-(\ref{kappatc}). We note that there is quantitative difference but the qualitative similarity between results obtained using critical exponents of SBM  and that of critical exponents of 3-d Ising model. It has been argued that near the QCD critical point, the bulk viscosity may show singular behaviour as $\zeta\sim |t|^{-a_{\zeta}}$, where $a_{\zeta}\sim1.8$\cite{onuki1997dynamic}. Similarly, the thermal conductivity may show singular behaviour $\kappa\sim|t|^{-a_{\kappa}}$, where $a_{\kappa}\sim 0.63$. Thus, the singular part of Rayleigh peak may behaves as,
  
  \begin{equation}
  \Gamma_T\sim|t|^{\hat{\gamma}-\alpha_{\kappa}}\nonumber.
  \end{equation}
  
 With the critical exponents of 3-d Ising model: $\hat{\gamma}=1.2$ and $\alpha_{\kappa}=0.63$, we get $\Gamma_{T}\sim |t|^{0.57}$.  In our case, we have

 \begin{equation}
  \Gamma_T\sim|t|^{\hat{\beta}-\hat{\nu}+3/2\hat{\alpha}}\nonumber.
  \end{equation} 
  
  Using critical exponents of SBM, we get $\Gamma_{T}\sim |t|^{0.25}$. This quantitative difference and qualitative similarity can also be seen in Brillouin peak. Using Ising critical exponents of 3-d Ising model we get $\Gamma_{S}\sim |t|^{-a_{\zeta}}=|t|^{-1.8}$,  whereas in our case $\Gamma_{S}\sim |t|^{-1.25}$.  This qualitative similarity and quantitative differences could be attributed to the value of specific heat critical exponent $\hat{\alpha}$. In case of universality class of 3-d Ising model $\hat{\alpha}\sim 0.11$ which is positive, and it is largely responsible for the rapid rise (in fact, divergence) of the bulk viscosity near the QCD critical point. In case of SBM, $\alpha=\frac{1}{2}$ which is again positive. So it turns out that the qualitative behaviour of the bulk viscosity and hence the correlation function is not different from that of 3-d Ising model universality class.

Let us discuss the implications of these results in the search for CP in heavy-ion collision experiments. Near the critical point the width and the strength of these modes depends on the singularities in static thermodynamic quantities and the transport coefficients. The critical exponents calculated within ambit of SBM are given by Eq.(\ref{ce}), while the dynamical critical exponents can be read out from (\ref{visce}). We note that both the bulk viscosity ($\zeta$) and thermal conductivity ($\kappa$) rise very rapidly near the critical point. Such rapid rise in the bulk viscosity has been estimated in previous studies\cite{Karsch:2007jc,Antoniou:2016ikh,Kadam:2020utt}. This rapid rise in both $\zeta$ and $\kappa$ render divergence in the width of the Brillouin peaks. However, as we have seen above, the strengths of the Rayleigh and  Brillouin peaks is governed by the ratio of two specific heats $\gamma=\frac{\tilde{C}_P}{\tilde{C}_{n}}$ which diverge as $T\rightarrow T_c$. Thus, the acoustic mode loose its strength near the critical point and  the divergent behaviour of the bulk viscosity may not be observed in the density correlations. 

One of the phenomenological consequence of the disappearance of sound mode near the critical point may be seen in the three-particle correlation due to the Mach-cone formation in the HIC. Such three particle correlation has been observed in the RHIC experiment\cite{Ploskon:2007es,PHENIX:2005zfm,Ayala:2011zza,STAR:2008kud}. If such three particle correlations are indeed due to Mach-cone formation, then the suppression or its disappearance altogether would signal the existence of the critical point. In this context a study carried out in Ref. \cite{Kapusta:2012zb} is  very interesting. This this work, authors have studied the effect of critical singularity in the thermal conductivity on two-particle correlation in the boost-invariant hydrodynamical model. It is found that the growth of the thermal conductivity near the critical point implies the existence of two-particle correlations over 2 units of rapidity. The strength of this correlation increases as the trajectory pass close to the critical point. One can carry out similar studies with the three-particle correlations related to Mach-cone formation in which the critical singularities in both the bulk viscosity and thermal conductivity are included the hydrodynamical simulations.

The critical singularities observed in the transport coefficients have some other phenomenological consequences as well. In Ref.\cite{Monnai:2016kud}, authors have discussed the effect of enhanced bulk viscosity near the critical point on the bulk hydrodynamical evolution of the matter created in heavy-ion collision. They incorporated critical expected behaviour of the bulk viscosity of the dynamical universality class of model H, namely $\zeta\sim |t|^{-2}$, within non-boost-invariant, longitudinally expanding $1+1$ dimensional causal relativistic hydrodynamical evolution at non-zero baryon density. They found, at forward rapidity, a sizeable increase in the quantities, $dN_{B}/dY$ and $dN_{ch}/dY$, where $N_{B}, N_{ch}$ and $Y$ respectively corresponds to the net-baryon multiplicity, charged particle multiplicity and momentum rapidity. In our case, since $\zeta\sim|t|^{-5/4}$ we may expect the similar conclusion. However, the deviation from non-CP behaviour won't be as sizeable as that of observed in \cite{Monnai:2016kud}. Since the thermal conductivity also behaves, in our model, as $\kappa\sim|t|^{-5/4}$, we expect that $dN_{B}/dY$ and $dN_{ch}/dY$ will be strongly affected. Rapid growth of bulk viscosity will also leads to softening of effective pressure which, in term, can lead to non-monotonic behaviour in the slope of the directed flow of net protons or that in the triangular flow.

\section{Summary and conclusion}
\label{secV}

To summarise, we have calculated the density correlations in a baryon rich relativistic fluid from Mori-Zwanzig-Nakajima projection operator formalism. The hydrodynamic form of density correlation can be obtained, with a judicious choice of slow variables, using this formalism. The spectral function of density correlation is found to consist of usual three peaks: two acoustic modes at $\omega=\pm C_s k$ and one thermal mode at $\omega=0$. The width of the Rayleigh peak is found to be the same as that in the non-relativistic case, whereas the width of Brillouin peak consist of extra relativistic correction which is negative and hence reduces its width. 

We further discussed the behaviour of Rayleigh and Brillouin peaks near the critical point. We extracted the singular behaviour of thermodynamic quantities using statistical bootstrap model. The singular behaviour of transport coefficient, shear and bulk viscosity as well as thermal conductivity can be extracted using the ansatz for this quantities given by Eq.(\ref{ansatz}). We found that the bulk viscosity contributes to the sound attenuation at leading order  $\sim|t|^{-\frac{5}{4}}$ while the thermal conductivity contribute at sub-leading order $\sim|t|^{-\frac{3}{4}}$. On the other hand, only the thermal conductivity contributes at leading order $\sim|t|^{\frac{1}{4}}$ to the thermal diffusivity. We noted that the acoustic mode loose its strength near the critical point and  the divergent behaviour of the bulk viscosity may not be observed in the density correlations. However, it can be inferred from the average transverse momenta and multiplicities of produced particles in heavy ion collisions. 

It is important to note that SBM is just an effective model of QCD. The critical exponents extracted within ambit of SBM are different from that of 3-d Ising model. One of the  reason for this discrepancy would be that SBM does not take into account effective glueball degrees of freedom.  It would be interesting to explore this extension of SBM and study the critical properties within its ambit. Further, we have completely ignored the non-linear terms in Eqs. (\ref{landau4})-(\ref{landau6}) which plays an important role in determining the dynamical critical exponents\cite{Koide:2004ph}. Inclusion of such terms would be an interesting improvement in our calculations. Work in these directions is under progress and will appear somewhere else.
 
In conclusion, density correlations may not be suitable observable to search for the QCD critical point. However, as noted in Ref.\cite{Minami:2009hn} suppression or disappearance of Mach cone, which is related to the existence of sound mode in a fluid, may indicate that system have passed close to the critical point.

\appendix
\label{appendix1}
\section{Calculation of the memory matrices}
We shall briefly discuss the calculation of the memory matrices. The details can be found in Refs.\cite{mazenko2008nonequilibrium,Minami:2012hs}.
\subsection{Streaming matrix}
The streaming matrix can be written as,
\begin{gather}
i{\bf{K}}^{(s)}\equiv\begin{bmatrix} 
	K^{(s),n}_{n} & K^{(s),e}_{n} & K^{(s),{\bf{p}}}_{n} \\
	K^{(s),n}_{\epsilon} & K^{(s),\epsilon}_{\epsilon} & K^{(s),{\bf{p}}}_{\epsilon}\\
	K^{(s),n}_{\bf{p}} & K^{(s),\epsilon}_{\bf{p}} & K^{(s),{\bf{p}}}_{\bf{p}} \\
	\end{bmatrix}
\end{gather}
 where ${\bf{p}}=h_0{\bf{v}}$. From Eq.(\ref{smf}) we can write,

\begin{eqnarray}
\label{ks11}
iK_{{n}_b}^{(s){n}_b}({\bf{k}})&=&(i\mathcal{L}\delta\hat{n}_b({\bf{k}}), \mathcal{\hat{A}}^{{n}_b}(0))\\\nonumber
&=&(i\mathcal{L}\delta\hat{n}_b({\bf{k}}), \mathcal{\hat{A}}_{l}(0))g^{l{n}_b}({\bf{k}})\\\nonumber
&=&(i\mathcal{L}\delta\hat{n}_b({\bf{k}}), \delta\hat{n}_{b}(0))g^{{n}_b{n}_b}({\bf{k}})+(i\mathcal{L}\delta\hat{n}_b({\bf{k}}), \delta\hat{\epsilon}(0))g^{\epsilon n_b}({\bf{k}})\\\nonumber
&+&(i\mathcal{L}\delta\hat{n}_b({\bf{k}}), \delta\hat{{\bf{p}}}(0))g^{{\bf{p}}n_b}({\bf{k}})\\\nonumber
&=&-ik^{l}\big[g_{{j}^{l}_b{n}_b}({\bf{k}})g^{{n}_b{n}_b}({\bf{k}})+g_{j^{l}_{b}\epsilon}({\bf{k}})g^{\epsilon n_{b}}({\bf{k}})+g_{j^{l}_{b}p^{j}}({\bf{k}})g^{p^{j}n_b}({\bf{k}})\big]
\end{eqnarray}

 where $j_{b}$ is the baryon current. If we assume that the equilibrium density matrix $\hat{\rho}_0$ is time-translational invariant then $\delta\hat{n}_b$ and $\delta\hat{\epsilon}$ are even functions of time while $\delta\hat{p}^{i}$ and $\delta \hat{j}_{b}^{i}$ are odd.  Hence the time reversal symmetry implies that $\delta\hat{n}_b$ and $\delta\hat{T}$ do not mix with $\delta\hat{p}^{i}$ and $\delta \hat{j}_{b}^{i}$. Thus Eq.(\ref{ks11}) implies that $iK_{{n}_b}^{(s){n}_b}=0$. This further implies that $iK_{{n}_b}^{(s)\epsilon}$ also vanishes and only $iK_{{n}_b}^{(s){\bf{p}}}({\bf{k}})$ survives.

\begin{eqnarray}
 iK_{{n}_b}^{(s)p^i}({\bf{k}})&=&(i\mathcal{L}\delta\hat{n}_b({\bf{k}}),\delta\hat{v}_{l}(0))g^{p^{l}p^{i}}({\bf{k}})\\\nonumber
 &=&-ik^{j}(\delta \hat{j}^{j}_b({\bf{k}}),\delta\hat{v}^{l}(0))g^{p^{l}p^{i}}({\bf{k}})\\\nonumber
 &=&-ik^{j}  g_{j_{b}^{j}p^{l}}({\bf{k}}) g^{p^{l}p^{i}}({\bf{k}})
 \label{ks131}
\end{eqnarray}

In the low energy limit ${\bf{k}}\rightarrow 0$ we can expand $g_{mn}({\bf{k}})$ around ${\bf{k}}=0$ as a power series as,

\begin{equation}
g_{mn}({\bf{k}})=g_{mn}({\bf{0}})+{\bf{k}}\cdot\nabla_{{\bf{k}}}g_{mn}({\bf{k}})+\mathcal{O}(k^2)
\label{metricexp}
\end{equation}

At linear order only the leading term $g_{mn}({\bf{0}})$ contributes in (\ref{metricexp}). Further, $g_{mn}({\bf{0}})$ satisfy following relations\cite{}:

\begin{equation}
g_{p^{i}p^{j}}({\bf{0}})=\int d^3x (\hat{T}^{0i}(0,{\bf{x}}),\hat{T}^{0j}(0,{\bf{0}}))=\delta^{ij}T_0h_0
\label{gpp}
\end{equation}

\begin{equation}
g_{p^{i}j_{b}^{j}}({\bf{0}})=\int d^3x (\hat{T}^{0i}(0,{\bf{x}}),\hat{j}_{b}^{j}(0,{\bf{0}}))=\delta^{ij}T_0n_0
\label{gpj}
\end{equation}

Using Eqs.(\ref{gpp}) and (\ref{gpj}), Eq(\ref{ks131}) reduces to

\begin{equation}
 iK_{{n}_b}^{(s)p^i}({\bf{k}})\simeq-ik^iT_0n_0
\end{equation}

We can similarly calculate $ iK_{\epsilon}^{(s)m}$ and $iK_{v^i}^{(s)m}$. The streaming matrix is finally written as

\begin{gather}
i{\bf{K}}^{(s)}\equiv\begin{bmatrix} 
	0 & 0 & -ik^in_0T_0 \\
	0 & 0 & -ik^i\\
	-ik^in_0T_0 & -ik^in_0T_0 & 0 \\
	\end{bmatrix}
	\label{smm}
\end{gather}

\subsection{Dynamic memory matrix}

The dynamic memory matrix can be written as,
\begin{gather}
i{\bf{K}}^{(d)}\equiv\begin{bmatrix} 
	K^{(d),n}_{n} & K^{(d),\epsilon}_{n} & K^{(d),{\bf{p}}}_{n} \\
	K^{(d),n}_{\epsilon} & K^{(d),\epsilon}_{\epsilon} & K^{(d),{\bf{p}}}_{\epsilon}\\
	K^{(d),n}_{\bf{p}} & K^{(d),\epsilon}_{\bf{p}} & K^{(d),{\bf{p}}}_{\bf{p}} \\
	\end{bmatrix}
\end{gather}

 From Eq. (\ref{dmf}) we write

\begin{equation}
K_{n_b}^{(d)m}(t-s,{\bf{k}})\equiv-\theta(t-s) \big(i\mathcal{\hat{L}}e^{it\mathcal{\hat{Q}}\mathcal{\hat{L}}}\:\mathcal{\hat{Q}}\:i\mathcal{\hat{L}}\delta\hat{n}_b(0,{\bf{k}}), \mathcal{\hat{A}}^{m}(s,{\bf{0}})\big)
\end{equation}

In this case, since $\delta\hat{{\bf{p}}}$ is a slow variable there will not be dissipative terms in the generalized Langevin equation for $\delta\hat{\epsilon}$. Hence we get $i\hat{\mathcal{L}}\delta\hat{\epsilon}=-{\bf{k}}\cdot \delta\hat{{\bf{p}}}$. Further, since $\delta\hat{{\bf{p}}}$ is a slow variable, its orthogonal projection should vanish. Thus $\hat{\mathcal{Q}}\:i\hat{\mathcal{L}}\delta\hat{\epsilon}=0$ and hence $K_{n_b}^{(d)\epsilon}$ vanishes. Further, $K_{n_b}^{(d){\bf{p}}}\simeq \mathcal{O}(k^3)$, which we neglect.  Thus, only only $K_{n_b}^{(d)n_b}$ survives. It can be easily calculated in the Markov approximation. The final expression is written as:

\begin{equation}
K_{n_b}^{(d)n_b}({\bf{k}})\simeq{\bf{k^2}}\bigg(\frac{n_0T_0}{h_0}\bigg)^2\kappa
\end{equation}

where we have define the thermal conductivity coefficient as,
\begin{equation}
\kappa=\bigg(\frac{h_0}{n_0T_0}\bigg)^2\int_0^\infty dt \int d^3z \bigg(\delta {\bf{j}_{q}}(t,{\bf{x}}),\delta{\bf{j}_{q}}(0,{\bf{0}})\bigg)
\end{equation}

with the heat current defined as $\delta {\bf{j}_{q}}\equiv\delta {\bf{j}_{b}}-\frac{n_0}{h_0}\delta {\bf{p}}$. One can similarly carry out the calculation of other elements in the $i{\bf{K}}^{(d)}$. The final result is:

\begin{gather}
i{\bf{K}}^{(d)}\equiv\begin{bmatrix} 
	{\bf{k^2}}\tilde\kappa & 0 & \mathcal{O}(k^3) \\
	K^{(d),n}_{\epsilon} & K^{(d),\epsilon}_{\epsilon} & K^{(d),{\bf{p}}}_{\epsilon}\\
	\mathcal{O}(k^3) & 0 & T_0\tilde\nu k^ik^j+T_0\eta {\bf{k}}^2\delta^{ij} \\
	\end{bmatrix}
\end{gather}

where  $\tilde\kappa=\bigg(\frac{n_0T_0}{h_0}\bigg)^2\kappa$ and $\tilde\nu=\bigg(\zeta+\frac{1}{3}\eta\bigg)$. $\kappa$ is the thermal conductivity, $\eta$ is the shear viscosity and $\zeta$ is the bulk viscosity.

\vspace{5mm}
\section*{Acknowledgments}
Guruprasad~Kadam is financially supported by the DST-INSPIRE faculty award under Grant No. DST/INSPIRE/04/2017/002293. G.K. thanks Amaresh Jaiswal for useful discussion.

\bibliography{projection}

\end{document}